\documentclass[twocolumn,showpacs,prl,10pt]{revtex4}
\usepackage[english]{babel}
\usepackage{amsmath,amssymb}
\usepackage{times}
\usepackage{epsfig}

\begin{document} 
\title{Breakdown of the Luttinger sum rule within the Mott-Hubbard
  insulator}

\author{J. Kokalj$^{1}$ and P. Prelov\v sek$^{1,2}$}
\affiliation{$^1$J.\ Stefan Institute, SI-1000 Ljubljana, Slovenia}
\affiliation{$^2$ Faculty of Mathematics and Physics, University of
Ljubljana, SI-1000 Ljubljana, Slovenia}

\date{\today}
                   
\begin{abstract}
The validity of the Luttinger sum rule is investigated within the
prototype tight-binding model of interacting fermions in one
dimension, i.e., the $t$-$V$ model including the next-nearest neighbor hopping $t'$
in order to break the particle-hole symmetry. Scaling analysis of
finite-system results at half-filling reveals evident breakdown of the
sum rule in the regime of large gap at $V \gg t$, while the sum rule
appears to recover together with vanishing of the Mott-Hubbard gap.
\end{abstract}

\maketitle
The Luttinger theorem \cite{luttinger60b,luttinger60a} is the
essential building block supporting the concept of the Fermi liquid
(FL) as formulated by Landau \cite{landau56}. The Luttinger sum rule
(LSR) in a homogeneous system relates the Fermi volume to the density
of fermions irrespective of the presence of the electron-electron
interactions.  Stimulated by experiments on novel electronic materials
with strongly correlated electrons which indicate possible deviations
from FL scenario and from LSR \cite{yoshida07}, theoretical studies
of the validity of the LSR and its limitations have intensified.

In a metal at $T=0$ the Fermi surface is located by the poles of the
Green's function (GF) $G({\bf k},\omega=0)$ and the LSR shows that the
Fermi volume being equal to the density of electrons $n$ is unchanged
by the interaction. It has been pointed out
\cite{dzyalosh03,altshuler98} 
that the original derivation \cite{luttinger60b} and the LSR can be
generalized as well to insulators where the corresponding 'Luttinger
surface' (LS) is defined by zeros $G({\bf k},0)=0$. Such LSR concept
becomes of interest, but as well easier to test in strongly interacting
electrons and in Mott-Hubbard (MH) insulators in particular. It has
been recently applied to spin ladders \cite{konik06}. On the other
hand, there are several indications that LSR might be violated
within the MH insulators in general \cite{stanescu07,rosch07,gros05}.  The
argument is based on the observation, that the LSR is satisfied only for a
particular value of chemical potential $\mu$ within the MH gap. It has
been shown that for models with the particle-hole (p-h) symmetry the
latter is the case and the LSR is fulfilled  \cite{stanescu07}. At the
same time, it has 
been realized that the LSR should as well apply to finite systems
\cite{kokalj07,ortloff07}. This allows us to test validity of LSR in nontrivial
models of correlated electrons \cite{kokalj08}. Based on analytical
expansion for $U/t \gg 1$ it has been shown on small systems that
within Hubbard model on a planar triangular lattice (without the p-h
symmetry)  LSR is indeed violated for a range of parameters
\cite{kokalj08}.

In this Letter we present results of the numerical study within the
prototype model of interacting fermions in 1D, i.e., the generalized
$t$-$V$ model. The advantage of such a 1D model is
that it allows for the finite-size scaling to the thermodynamic limit. Our
results show clear violation of the LSR for $V \gg t$ within MH
insulating phase which appears to persists down to critical $V>V_c$
where the MH gap opens.

In the following we study the extended $t$-$V$ model
\begin{eqnarray}
H=&-&t \sum_i (c_{i+1}^\dagger c_i + \textrm{h.c.}) -t' \sum_i
(c_{i+2}^\dagger c_i + \textrm{h.c.}) + \nonumber \\
&+& V\sum_i n_i n_{i+1}, \label{hamil}
\end{eqnarray}
where $t$ and $t'$ are nearest-neighbor (n.n.) and
next-nearest-neighbor (n.n.n.) hopping, respectively, and $V$ is the
n.n. repulsive interaction between fermions. We are interested in the
MH insulator state which appears at half-filling with the electron density
$n=1/2$. It is well known that the model with $t'=0$ is equivalent to
the anisotropic Heisenberg model, which can be solved exactly via
Bethe ansatz \cite{cloizeaux66}. The model shows the transition from
the metallic state 
$V<V_c$ to a MH insulator for $V>V_c$ with $V_c=2t$.  Our study is
focused on systems with non-zero $t'$, for which there is no
exact solution. We choose such a system due to the lack of p-h
symmetry, since for $t'=0$ the LSR is automatically satisfied for
$n=1/2$ \cite{stanescu07}. In the following we study $t'/t=0.4$, $0.2$ in order
to have substantial deviation from the p-h symmetry, but at the same time
to maintain the simple momentum distribution for noninteracting fermions.

\begin{figure}[htb]
\begin{center}
  \includegraphics[width =0.9\linewidth] {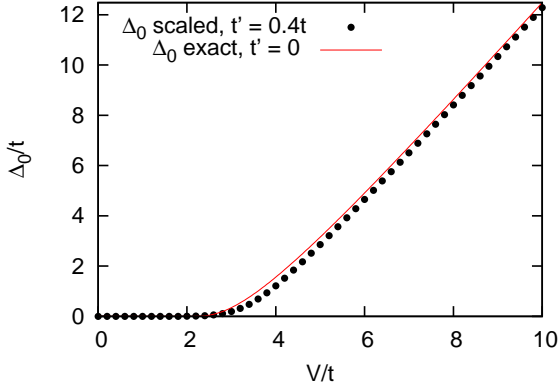} 
\end{center}
\caption{(Color online) Scaled gap $\Delta_0$ for non-symmetric case $t'=0.4t$ (points), and exact
Bethe ansatz results for $t'=0$ (line).}
\label{gapimg}
\end{figure}

Let us first consider the opening of the MH gap $\Delta_0$ at
$n=1/2$. It is well known \cite{cloizeaux66} that within the $t$-$V$ model
the gap opens at $V=2t$ being exponentially small for $V >2t+ $ and
nearly linear in $V$ for $V>4t$, see Fig.~1. We analyze the effect of
$t'$ on $\Delta_0$  by performing the 
exact diagonalization of chains with $N=14, 18, 22, 26,$ and $30$ sites
using the Lanczos technique. Chains (with periodic boundary
conditions) were chosen to have odd number of electrons $N_e=N/2$,
since this leads to a non-degenerate ground state. The MH gap is then
determined via
\begin{equation}
\Delta(N)=(E_0^{N_e+1} -E_0^{N_e})-  (E_0^{N_e} -  E_0^{N_e-1}).
\end{equation}
To obtain the gap in thermodynamic limit as $\Delta_0 = \lim_{N\to
\infty} \Delta(N)$ we perform the finite-size scaling for $N=14 -30$
using $\Delta(N)= a + b/N+ c/N^2$. In the metallic regime, $V<2t$, the
gap scales expectedly with $a \sim 0$, and at large $V\gg 2t $ with $b
\sim 0$.  In general parameters $a,b,c$ were obtained from least-square
fit. Scaled gap $\Delta_0=a $ for $t'=0.4t$ is shown in
Fig.~\ref{gapimg}, together with the exact Bethe ansatz \cite{cloizeaux66}
result for $t'=0$. The result reveals that the gap for non-symmetric case
deviates only slightly from the exact result for symmetric case.

The (retarded) Green's function 
$G(k,\omega)$  at $T=0$ is defined as  
\begin{eqnarray}
G({k},\omega)=-i\int_{0}^\infty dt \mathrm{e}^{i(\omega+\mu)t}
\langle0| \{c_{{ k}}^\dagger, c_{{ k}}(t)\}_+ |0\rangle  \nonumber,
\label{gf1}
\end{eqnarray}
where $\mu$ is the chemical potential. We are studying finite systems
at fixed number of electrons, $N_e=N/2$. Clearly, the position of
$\mu$ within the MH gap is crucial for further discussion of LSR
\cite{rosch07,stanescu07}. Within this approach the correct choice is
\cite{perdew82,kaplan06}
\begin{equation}
\mu(N)=(E_0^{N_e+1}-E_0^{N_e-1})/2.
\label{mu}
\end{equation}

\begin{figure}[htb]
\begin{center}
  \includegraphics[width =0.8\linewidth] {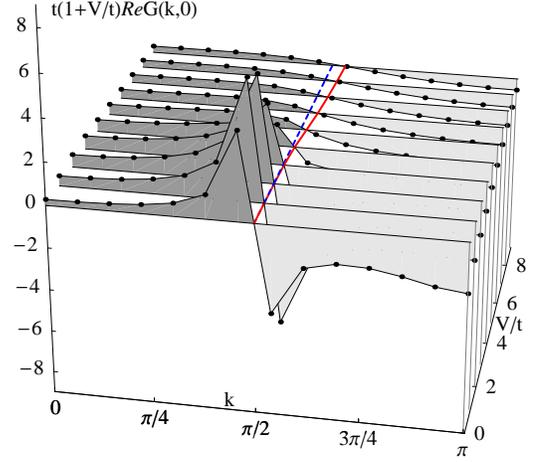} 
\end{center}
\caption{(Color online) Evolution of  ${\mathrm Re}G(k,0)$ with increasing interaction
$V$ for $t'/t=0.4$ and for half-filled system of $N=26$ sites with allowed wave vectors
  (points) and indications of $k_F=\pi/2$ (dashed line) and $k_L$
  (full line). Points are connected with straight lines. 
}
\label{gfplot}
\end{figure}

For LSR ${\mathrm Re} G(k,0)$ is important.  In order to start with a
general perspective we present in Fig.~\ref{gfplot} the evolution of
$G(k,0)$ with increasing $V$.  For non-interacting system
${\mathrm Re}G(k,0)$ has a singularity at $k_F=\pi/2$ persisting apparently
within the metallic state, $V<2t$. This is the normal Fermi-liquid (in 1D
Luttinger-liquid) behavior connected with the poles (singularities) of the spectral
function $A(k,\omega)=-{\mathrm Im}G(k,\omega)/\pi$ approaching
$\omega=0$ for $k \to k_F$.  With increasing $V/t$ and the opening of
the MH gap the behavior 
changes qualitatively. Within the gap $A(k,\omega)=0$, hence $G(k,0)$
has no singularity and goes through zero smoothly at $k \sim \pi/2$.
Moreover, for $V>4t$, $G(k,0)$ becomes small for all $k$. This
happens because at $V\gg t$, $A(k,\omega)$ consists of two nearly
equal weights at approximately $\pm V$ and their contributions to real
part of GF almost cancel each other at the chemical potential, therefore
$G(k,0)\sim 1/V^2$ \cite{kokalj08}. 

Let us now focus on the LSR and its breakdown. The content of LSR is the
precise locus of $k=k_L$, where $G(k,0)$ changes sign
\cite{luttinger60a,luttinger60b}, $k_L$ called Luttinger wave vectors
\cite{dzyalosh03}. According to the LSR, for spinless model at $n=1/2$ one should
generally have $k_L=\pi/2$, if the topology of the electronic band is
not changed qualitatively (which could happen, e.g. for $t' >0.5t$).
From Fig.~\ref{gfplot} we note that $k_L$ is indeed near
$\pi/2$, however, even without finite-size scaling a small deviation
$k_L \ne \pi/2$ may be observed for $V>4t$. More accurate analysis
with the finite-size scaling is presented below. 

To determine $k_L$ in the thermodynamic limit $N \to \infty$, we perform
the finite-size scaling of results at various $N$. Our procedure is
the following. Due to periodic boundary conditions, allowed are $k=2
\pi l/N$. Since we work with nondegenerate case with odd $N_e=N/2$,
$k=\pi/2$ does not appear directly for any system. The closest are,
however, $k^-(N)=\pi/2 - \pi /N$ and $k^+(N)=\pi/2 + \pi /N$. Next we
determine $\mu$ using Eq.(\ref{mu}) for given $N$. Then, the GF is evaluated via ED
using the Lanczos algorithm for $T=0$ dynamical quantities
\cite{haydock72} , i.e., from 
Eq.(\ref{gf1}), we have $G=G^c + G^a$,
\begin{equation}
G^c(k,\omega)= \langle 0|c_k^\dagger (\omega+\mu+E_N^0 -H)^{-1} c_k
|0\rangle,
\end{equation}
and analogous for $G^a$.  In particular, we calculate values of
GF at $k^\pm(N)$,
\begin{equation}
G^\pm(N)=G_N(k^\pm(N),0),
\end{equation}
where $G_N(k,\omega)$ stands for the GF of a system-size $N$.
These values are used to calculate $k_L$ in the following manner. For
each $N$, we evaluate the mean value $\bar G (N)=
(G^+(N)+G^-(N))/2$ and the difference $\Delta G (N)= G^+(N)-G^-(N)$,
which lateron serve for the evaluation of the derivative $\partial
G(k,0)/\partial k$. Next, we perform scaling of both, $\bar G(N)$ and
$\Delta G(N)$, to obtain their values in the limit $N\to \infty$.

\begin{figure}[htb]
\begin{center}
  \includegraphics[width =0.9\linewidth] {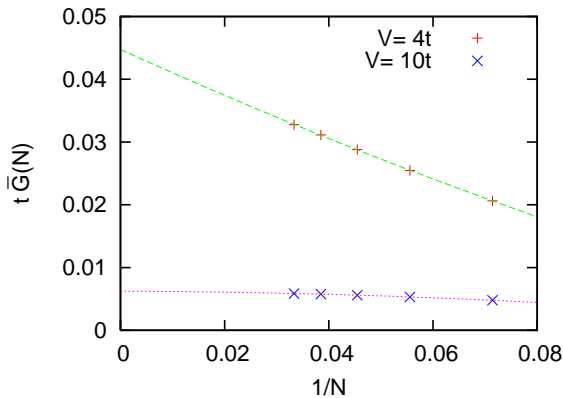} 
\end{center}
\caption{(Color online) Scaling of $\bar G(N)$ for two values of $V/t$ and $t'=0.4t$.
}
\label{bargplot}
\end{figure}

Finite-size scaling of $\bar G(N)$ is performed by assuming 
\begin{equation}
\bar G(N)=a_1+b_1\frac{1}{N}+c_1\frac{1}{N^2}.
\label{barg}
\end{equation}
Quadratic term $c_1\frac{1}{N^2}$ is included in analogy with
$\Delta(N)$. For $V\sim 4t$ dominant is the linear term $1/N$, but
with increasing $V$, parameter $b_1$ in Eq. (\ref{barg}) decreases and
$c_1$ term becomes more important. In Fig. \ref{bargplot} values of
$\bar G(N)$ and obtained scaling from least square fit is shown for
$t'/t = 0.4$ and for
two values $V/t= 4, 10$.  As it is seen from Fig.~\ref{bargplot}, the
relevant limiting value is $\bar G= \lim_{N\to \infty}\bar G(N)= a_1$.
Note that $a_1=G(\pi/2,0)$ should be zero according to the LSR. Obviously our
finding in Fig.~\ref{bargplot} that $a_1 \ne 0$ is the indication that
the LSR is violated.

\begin{figure}[htb]
\begin{center}
  \includegraphics[width =0.9\linewidth] {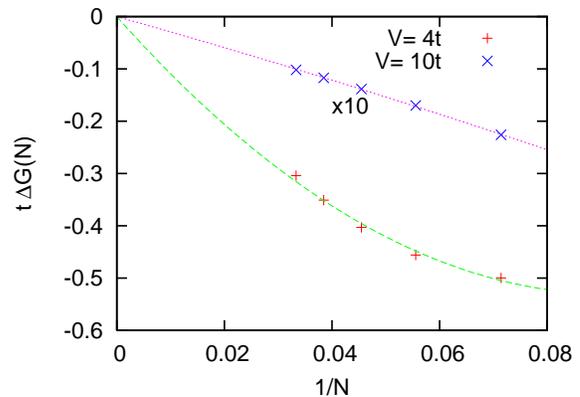} 
\end{center}
\caption{(Color online) Scaling of $\Delta G(N)$ for two values of $V/t$ and $t'=0.4t$.}
\label{tildegplot}
\end{figure}

For the estimate of $k_L$ also the scaling of $\Delta G(N)$ is needed.
$G(k,0)$ within an insulator is a continuous function of $k$, hence
$\Delta G(N)$ goes to zero as $N\to \infty$, so that the proper
scaling function is $\Delta G(N)= b_2/N+ c_2/N^2$. Least-squares fits
for the same parameters as in Fig. ~\ref{bargplot} are
shown in Fig. \ref{tildegplot}. We notice, that for large $V \gtrsim 10t$ the
linear term $b_2/N$ is dominant, while for $V\sim 4t$ quadratic
correction with  $c_2$ also becomes relevant.   

For further analysis the derivative of GF at $\pi/2$ is relevant 
\begin{equation}
\frac{\partial G(k,0)}{\partial k}|_{k=\pi/2}=\lim_{N\to \infty}
\frac{\Delta G(N)}{2\pi/N}=\frac{b_2}{2\pi}.
\end{equation}

To calculate the Luttinger momentum $k_L$ for which GF changes sign we
use linear approximation of GF near $\pi/2$ assuming that $k_L$ does
not deviate appreciably from $\pi/2$ being indeed the case. The result for
$k_L$ may in our approximation be written as
\begin{equation}
k_L=\frac{\pi}{2}-\frac{G(\pi/2,0)}{\partial G(k,0)/\partial
k|_{k=\pi/2}}=\frac{\pi}{2}-2\pi\frac{a_1}{b_2}.
\label{k_L}
\end{equation}

\begin{figure}[htb]
\begin{center}
  \includegraphics[width =0.9\linewidth] {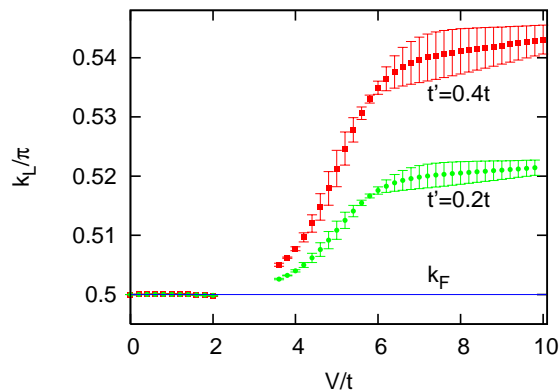} 
\end{center}
\caption{(Color online) Calculated Luttinger momentum $k_L$ vs. $V/t$ for two values
  of $t'/t$.  }
\label{klplot}
\end{figure}
Final results for two values $t'/t=0.2$, $0.4$, and within the whole
range of parameters $V/t$ are shown in Fig.~\ref{klplot} with
corresponding error bars. Presented values of $k_L$ in regime
$V>3.6t$ are calculated as described above via Eq. (\ref{k_L}). In a
window $2<V/t<3.6$ results are not shown since finite-chain
calculations become unreliable due to limited wave-vector resolution
and deviations from the simple finite-size scaling behavior.  On the
other hand, within the metallic regime $V<2t$, $G(k,0)$ has
singularity near $\pi/2$. Therefore it makes more sense to apply
instead a similar scaling analysis to the inverse values of GF,
$({\mathrm Re}G(k,0))^{-1}$, and to locate in this way $k_L$. Results
obtained in this way are shown in Fig. \ref{klplot} for regime $V<2t$.
Our estimate of error bar in Fig.~\ref{klplot} is given as the larger
value obtained either from standard deviations of parameters $a_1$ and
$b_2$ or from the difference of the scaled value without taking into
account the smallest system. 

From Fig.~\ref{klplot} it is evident that at large $V/t>6$, $k_L$
substantially deviates from the LSR prediction $k_F=\pi/2$.  The
deviation $\Delta k_L=k_L-\pi/2$ from LSR saturates at large $V \gg
t$, whereby its value scales with the asymmetry  given by $t'/t$.
At the same time, with the decreasing $V/t \to 2$ and with vanishing
of the MH gap $\Delta_0$, also $\Delta k_L$ appears to vanish. In fact,
in the regime $2<V/t<6$ the LSR deviation $\Delta k_L$ seems
qualitatively to follow the variation of $\Delta_0$. This is in
accordance with our observation that inverse derivative ($-1/b_2$) has
a similar behavior to $\Delta_0$ as a function of $V$. For
large $V>6t$, both $b_2$ as well as $a_2$ behave as $\propto 1/V^2$,
hence $\Delta k_L$ approaches a constant value which could be
evaluated via the method of moment expansion \cite{kokalj08}.
On the other hand, the analysis of data within the metallic phase for
$V<2t$ does not show (within our accuracy) any deviation from the LSR
$k_F=\pi/2$. Hence, our results are consistent with previous
confirmations of the LSR in the metallic phase away from half
filling \cite{blagoev97,yamanaka97}.  

In conclusion, our results clearly show that the LSR is violated in
the Mott-Hubbard insulator  within the 1D generalized $t$-$V$
model where the p-h symmetry is broken via the introduction of the
n.n.n. hopping $t' \ne 0$. Although we concentrated only on two values
of $t'/t=0.2$, $0.4$, the behavior is quite generic whereby the violation
$\Delta k_L$ seems to scale with the p-h asymmetry, at least for
modest parameter $t'/t$.
It should be stressed that substantially larger $t'/t$ can
perturb qualitatively the topology of the noninteracting band and
$n_k$ which
makes the interpretation more difficult. An important finding is that
the violation $\Delta k_L$ scales as the MH gap $\Delta_0$ on
approaching the metallic transition at $V = V_c$, while within the
gapless (metallic) phase we do not find any evidence for the LSR
violation.

Discussing possible generality of our finding, we first stress several
advantages of the extended 1D $t$-$V$ model analyzed above. Being a
spinless model it allows for an accurate enough study of largest
systems using the ED (compared, e.g., to the Hubbard model or
multiband models), in our case up to $N=30$ sites. Large span of
sizes then allows for a reliable finite-size scaling and extrapolation
to $N\to \infty$.

Another important ingredient is the absence of a
phase with a long range order in the phase diagram at $n \sim 1/2$.
The latter is due to the 1D character of the model and supported by the exact
solution at $t'=0$ \cite{cloizeaux66} which is not strongly perturbed by
moderate $t'$. Thus we are dealing solely with the MH metal-insulator
transition at $V=V_c$ and with its influence on the LSR. Note that in
analogous $D>1$ models studied so far, e.g., the 2D Hubbard model on a
square lattice \cite{kokalj07,stanescu07} or triangular lattice
\cite{kokalj08} as well 
as more general MH insulators \cite{rosch07}, the MH insulator is mostly accompanied
with an onset of a long-range magnetic ordering, while doped MH
insulators can show ferromagnetic order etc. An appearance of the
long-range order clearly spoils the translational invariance and thus
the validity of the LSR \cite{altshuler98} or at least requires the
reformulation of the latter.

In view of above discussion, the studied model is just the prototype
example of a model with the MH transition and one can expect that in
analogy the LSR would be generally violated within the MH insulators
without the p-h symmetry, even more if the MH transition is followed
with an ordered state breaking the translational or some other
symmetry. 

Quite open question is, however, the origin of the breakdown of the LSR,
or more precisely which part of the original proof
\cite{luttinger60b,luttinger60a,dzyalosh03} becomes invalid within the
MH insulator.  The 
basic argument \cite{luttinger60b} invokes the existence of the functional
constructed via the perturbation expansion in the interaction
strength. The required adiabatic connectivity \cite{altshuler98,rosch07,stanescu07} to
noninteracting fermions can be clearly questioned at the
metal-insulator transition which at least represents a nonanalytic point
at $n=1/2$ in the extended $t$-$V$ model, even at $t'=0$
\cite{cloizeaux66}. Moreover, it is evident that zeroes $G(k,0)=0$ in the MH
gap at the same time require divergent self energy $\Sigma(k,0) \to \infty$
\cite{dzyalosh03,stanescu07} which enhances doubts on the existence of an
appropriate functional. The question is important in connection with
the emerging conserving approximations for strongly correlated systems
based on a construction of such a functional but also on
dynamical-mean-field methods based on approximations for
$\Sigma(k,\omega)$ \cite{stanescu06,kotliar06}. In any case, a deeper understanding of
the limitations of the validity of LSR is still missing and can provide
new insight in the physics of strongly correlated electrons.


\end{document}